\documentclass[varenna]{cimento}
\usepackage{graphicx}
\usepackage{amsmath,amssymb,amsfonts}
\usepackage{mathrsfs}
\def\hn{\hat{n}}
\def\bra{\langle}
\def\ket{\rangle}
\def\beq{\begin{equation}}
\def\beqno{\begin{equation}\nonumber}
\def\eeq{\end{equation}}
\def\a{\alpha}
\def\b{\beta}
\def\ve{\varepsilon}
\def\g{\gamma}

\def\s{\sigma}
\def\t{\rm{t}}
\newcommand\mum{\mu\rm{m}}
\newcommand\muK{\mu\rm{K}}
\newcommand\kHz{\rm{kHz}}
\def\li6{$^6$\rm{Li}}
\newcommand\cd{c^\dagger}
\newcommand\cdknus{c^\dagger_{\mathbf{k}\nu\sigma}}
\newcommand\cknus{c_{\mathbf{k}\nu\sigma}}
\newcommand\psid{\psi^\dagger}

\newcommand\psidb{\psi^\dagger_{\beta}}

\newcommand\psib{\psi_{\beta}}
\newcommand\psiknu{\phi_{\mathbf{k}\nu}}

\newcommand{\bk}{\mathbf{k}}

\newcommand{\br}{\mathbf{r}}
\newcommand{\bR}{\mathbf{R}}
\newcommand{\Dk}{\mathbf{q}}
\newcommand{\Dom}{\Omega}

\newcommand{\iomn}{i\omega_n}
\newcommand{\xik}{\xi_{\mathbf{k}}}
\newcommand{\kF}{\mathbf{k}_F}
\newcommand{\vF}{\mathbf{v}_F}
\newcommand{\Gk}{\Gamma_{\mathbf{k}}}
\newcommand{\zk}{Z_\bk}
\newcommand{\ek}{\varepsilon_{\mathbf{k}}}
\newcommand{\eF}{\varepsilon_F}

\newcommand{\eknu}{\varepsilon_{\mathbf{k}\nu}}

\newcommand{\enu}{\varepsilon_{\nu 0}}

\newcommand{\hO}{\hat{O}}

\newcommand{\Jaf}{J_{\rm{AF}}\,}
\newcommand{\smax}{s_{\rm{max}}}
\newcommand{\pd}{p_{\,2}}
\newcommand{\tstar}{T_F^\star}
\title{Condensed Matter Physics With Light And Atoms:\\
Strongly Correlated Cold Fermions in Optical Lattices.
}
\author{Antoine Georges}
\institute{Centre de Physique Th\'eorique,
Ecole Polytechnique,
91128 Palaiseau Cedex, France}
\shortauthor{A.~Georges}

\begin{document}

\maketitle

\medskip
\begin{center}
Lectures given at the Enrico Fermi Summer School on "Ultracold
Fermi Gases"\\
organized by M.~Inguscio, W.~Ketterle and C.~Salomon\\
(Varenna, Italy, June 2006)
\end{center}

\begin{abstract}
Various topics at the interface between condensed matter physics and the physics
of ultra-cold fermionic atoms in optical lattices are discussed.
The lectures start with basic considerations on energy scales, and on
the regimes in which a description by an effective Hubbard model
is valid. Qualitative ideas about the Mott transition are then presented, both
for bosons and fermions, as well as mean-field theories of this phenomenon.
Antiferromagnetism of the fermionic Hubbard model at
half-filling is briefly reviewed. The possibility that
interaction effects facilitate adiabatic cooling is discussed, and the importance of
using entropy as a thermometer is emphasized. Geometrical frustration of the lattice,
by suppressing spin long-range order, helps revealing genuine Mott physics and
exploring unconventional quantum magnetism.
The importance of measurement techniques to probe quasiparticle excitations
in cold fermionic systems is emphasized, and a recent proposal based on
stimulated Raman scattering briefly reviewed.
The unconventional nature of these excitations
in cuprate superconductors is emphasized.
\end{abstract}

\section{Introduction: a novel condensed matter physics.}

The remarkable recent advances in handling ultra-cold atomic gases have given
birth to a new field: condensed matter physics with light and atoms.
Artificial solids with unprecedented degree of controllability can be realized
by trapping bosonic or fermionic atoms in the periodic potential created
by interfering laser beams (for a recent review,
see Ref.~\cite{bloch_review_natphys_2005}, and other lectures in this volume).

Key issues in the physics of strongly correlated quantum systems
can be addressed from a new perspective in this context.
The observation of the Mott transition of bosons
in optical lattices~\cite{greiner_mott_nature_2002,
jaksch_lattice_prl_1998} and of the superfluidity of fermionic gases
(see e.g.~\cite{greiner_bec_mol_nature_2003,jochim_bec_mol_science_2003,
zwierlein_bec_mol_prl_2003,bourdel_bec_bcs_prl_2004})
have been important milestones in this respect,
as well as the recent imaging of Fermi
surfaces~\cite{kohl_fermisurface_prl_2005}.

To quote just a few of the many promising roads for research
with ultra-cold fermionic atoms in optical lattices, I would emphasize:

\begin{itemize}
\item the possibility of studying and hopefully understanding better
some outstanding open problems of condensed matter
physics, particularly in strongly correlated regimes,
such as high-temperature superconductivity and its interplay with Mott
localization.
\item the possibility of studying these systems in regimes which are
not usually reachable in condensed matter physics (e.g under
time-dependent perturbations bringing the system out of equilibrium), and
to do this within a highly controllable and clean setting
\item the possibility of ``engineering'' the many-body wave function
of large quantum systems by manipulating atoms individually or globally
\end{itemize}

The present lecture notes certainly do not aim at covering all these topics !
Rather, they represent an idiosyncratic choice reflecting the interests
of the author. Hopefully, they will contribute in a positive manner to
the rapidly developing dialogue between
condensed matter physics and the physics of ultra-cold atoms.
Finally, a warning and an apology: these are lecture notes and not
a review article. Even though I do quote some of the original work I refer to,
I have certainly omitted important relevant references, for which I
apologize in advance.

\section{Considerations on energy scales.}
\label{sec:scales}

In the context of optical lattices, it is convenient to express energies in units of
the {\it recoil energy}:
\beq
\nonumber
E_R=\frac{\hbar^2k_L^2}{2m}
\eeq
in which $k_L=2\pi/\lambda_L$ is the wavevector of the laser and $m$ the mass of the atoms.
This is typically of the order of a few micro-Kelvins (for a YAG laser with
$\lambda_L=1.06\mum$ and \li6 atoms, $E_R\simeq 1.4\muK$).
When venturing in the cold atoms community, condensed matter physicists who
usually express energy scales in Kelvins (or electron-Volts...!)
will need to remember that, in units of frequency:
\beqno
1\,\muK\,\simeq\,20.8\,\kHz
\eeq

\begin{figure}
\begin{center}
\includegraphics[width=7cm]{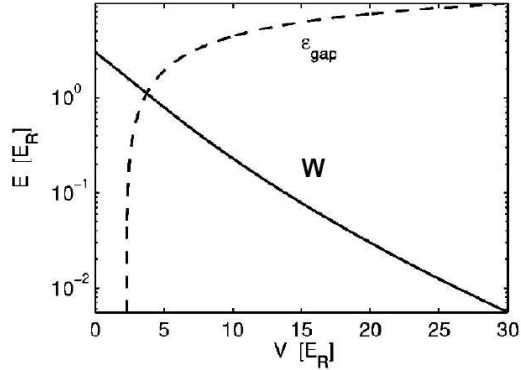}
\caption{Width of the lowest Bloch band and gap between the first two bands for
a 3-dimensional potential, as a function of laser intensity (in units of $E_R$)
(adapted from Ref.~\cite{blakie_cooling_fermions}). Note that in 3 dimensions, the
two lowest bands overlap for a weak lattice potential, and become separated only for
$V_0\gtrsim 2.3 E_R$.}
\label{fig:bandwidth}
\end{center}
\end{figure}

The natural scale for the kinetic energy (and Fermi energy) of atoms in the optical lattice
is not the recoil energy however, but rather the bandwidth $W$ of the Bloch
band under consideration, which strongly depends on the laser intensity $V_0$.
For a weak intensity $V_0\ll E_R$, the bandwidth $W$ of the lowest Bloch band in the
optical lattice is of order $E_R$
itself (the free space parabolic dispersion $\hbar^2 k^2/2m$ reaches the boundary of the
first Brillouin zone at $k=\pi/d=k_L$ with $d=\lambda_L/2$ the lattice spacing, so that
$W\simeq E_R$ for small $V_0/E_R$).
In contrast, for strong laser intensities, the bandwidth can be much smaller
than the recoil energy (Fig.~\ref{fig:bandwidth}). This is because in this limit
the motion of atoms in the lattice corresponds to tunneling between two neighboring potential
wells (lattice sites), and the hopping amplitude~\footnote{I could not force myself to use
the notation $J$ for the hopping amplitude in the lattice, as often done in the quantum
optics community. Indeed, $J$ is so commonly used in condensed matter physics to denote
the magnetic superexchange interaction that this can be confusing.
I therefore stick to the condensed matter notation $\t$, not to be confused of course
with time $t$, but it is usually clear from the context.}
$\t$ has the typical exponential dependence
of a tunnel process. Specifically, for a simple separable potential
in $D$ (=$1,2,3$) dimensions:
\beq
V(\br)= V_0\,\sum_{i=1}^D \sin^2 k_L r_i
\label{eq:sep_potential}
\eeq
one has~\cite{zwerger_mott_jopt_2003}:
\beq
\t/E_R = 4\pi^{-1/2} (V_0/E_R)^{3/4}\, e^{-2(V_0/E_R)^{1/2}}\,\,\,,
\,\,\,V_0\gg E_R
\label{eq:hop_deep}
\eeq
The dispersion of the lowest band is well approximated by a simple tight-binding expression
in this limit:
\beq
\ek=-2\t\sum_{i=1}^D \cos k_i
\eeq
corresponding to a bandwidth $W=4D\t\ll E_R$. The dependence
of the bandwidth,
and of the gap between the
first two bands, on $V_0/E_R$ are displayed on Fig.~\ref{fig:bandwidth}.

Since $W$ is much smaller than $E_R$ for deep lattices, one may worry that
cooling the gas into the degenerate regime might become very difficult.
For non-interacting atoms, this indeed requires $T\ll\eF$, with $\eF$ the
Fermi energy (energy of the highest occupied state), with $\eF\leq W$ for densities
such that only the lowest band is
partially occupied. Adiabatic cooling may however come to the rescue when
the lattice is gradually turned on~\cite{blakie_cooling_fermions}.
This can be understood from a very simple argument, observing that the
entropy of a non-interacting Fermi gas in the degenerate regime is limited
by the Pauli principle to have a linear dependence on temperature:
\beqno
S\,\propto T\,D(\eF)
\eeq
where $D(\varepsilon)$ is the density of states. Hence, $TD(\eF)$ is expected
to be conserved along constant entropy trajectories. $D(\eF)$ is inversely proportional
to the bandwidth $W$ (with a proportionality factor depending on the density, or
band filling): the density of states is enhanced considerably as the band shrinks
since the one-particle states all fit in a smaller and smaller energy window.
Thus, $T/W$ is expected to be essentially constant as the lattice
is adiabatically turned on: the degree of degeneracy is preserved and adiabatic
cooling is expected to take place. For more details on this
issue, see Ref.~\cite{blakie_cooling_fermions}
in which it is also shown that when the second band is populated, heating
can take place when the lattice is turned on (because of the increase of the
inter-band gap, cf. Fig.~\ref{fig:bandwidth}).
For other ideas about cooling and heating effects upon turning on
the lattice, see also Ref.~\cite{hofstetter_superfluidity_prl_2002}.
Interactions can significantly modify these effects, and
lead to additional mechanisms of adiabatic cooling, as discussed later in
these notes (Sec.~\ref{sec:cool}).

Finally, it is important to note that, in a strongly correlated system,
characteristic energy scales are in general
strongly modified by interaction effects in comparison to their bare, non-interacting values.
The effective mass of quasiparticle excitations, for example,
can become very large due to strong interaction effects, and correspondingly the scale
associated with kinetic energy may become very small. This will also be the scale below which coherent
quasiparticle excitations exist, and hence the effective scale for Fermi degeneracy.
Interaction effects may also help in adiabatically cooling the system however, as discussed later in
these notes.

\section{When do we have a Hubbard model ?}
\label{sec:hubbard}

I do not intend to review here in details the basic principles behind
the trapping and manipulation of cold atoms in optical lattices. Other lectures
at this school are covering this, and there are also excellent reviews
on the subject, see e.g
Refs.~\cite{bloch_review_natphys_2005,jaksch_toolbox,zwerger_mott_jopt_2003}.
I will instead only summarize the basic
principles behind the derivation of the effective hamiltonian. The focus
of this section will be to emphasize that there are some limits on the range
of parameters in which the effective hamiltonian takes the simple
single-band Hubbard form~\cite{werner_cooling_2005,werner_dea}.

\begin{figure}
\begin{center}
\includegraphics[width=6cm]{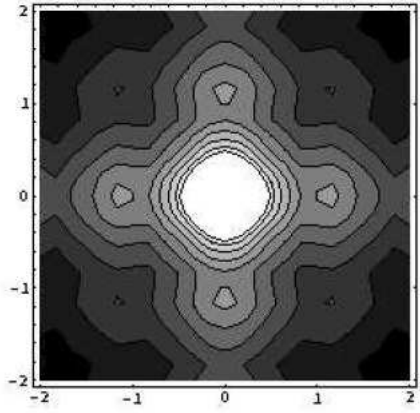}
\caption{Contour plot of the Wannier function corresponding to the lowest band
in the two-dimensional separable potential (\ref{eq:sep_potential}) with $V_0/E_R=10$.
The function has the symmetry of the square lattice, and has secondary maxima
on nearest-neighbor sites. The intensity of these secondary maxima control the
hopping amplitude. From Ref~\cite{werner_dea}.}
\label{fig:wannier}
\end{center}
\end{figure}

I consider two-component fermions (e.g two hyperfine states of identical
atomic species). The hamiltonian consists in a one-body term and an interaction term:
\beq
H=H_0 + H_{\rm{int}}
\eeq
Let me first discuss the one-body part, which involves the lattice potential
$V_L(\br)$ as well as the potential of the trap (or of the Gaussian waist of
the laser) $V_T(\br)$:
\beq
H_0 = \sum_\sigma \int d\br\,
\psi_\s^\dagger(\br)
\left[-\frac{\hbar^2\nabla^2}{2m} + V_L(\br) + V_T(\br) \right]
\psi_\s(\br) \equiv H_{0L} + H_{0T}
\eeq
The trapping potential having a shallow curvature as compared to the lattice
spacing, the standard procedure consists in finding first the Bloch states of
the periodic potential (e.g treating afterwards the trap in the local density
approximation). The Bloch functions $\psiknu(\br)$ (with $\nu$ an index labelling the band)
satisfy:
\beq
H_{0L}|\psiknu\ket=\eknu|\psiknu\ket
\eeq
with $\psiknu(\br)=e^{i\bk\cdot\br}u_{\bk\nu}(\br)$ and $u_{\bk\nu}$ a
function having the periodicity of the lattice. From the Bloch functions, one can construct
Wannier functions $w_{\bR\nu}(\br)=w_\nu(\br-\bR)$, which are localized around a specific
lattice site $\bR$:
\beq
w_{\bR\nu}(\br)=w_\nu(\br-\bR) = \sum_\bk e^{-i\bk\cdot\bR}\,\psiknu(\br)
=\sum_\bk e^{i\bk\cdot(\br-\bR)}\,u_{\bk\nu}(\br)
\eeq
In Fig.~\ref{fig:wannier}, I display a contour plot of the Wannier function corresponding to
the lowest band of the 2-dimensional potential (\ref{eq:sep_potential}).
The characteristic spatial extension of the Wannier function associated
with the lowest band is
$l_1\sim d$ (the lattice spacing itself) for a weak potential $V_0\ll E_R$,
while $l_1/d\sim (E_R/V_0)^{1/4}\ll 1$ for a deep lattice $V_0\gg E_R$.
The latter estimate is simply the extent of the ground-state wave-function of the
harmonic oscillator in the quadratic well approximating the bottom of the potential.

The fermion field operator can be decomposed on the localised Wannier functions basis set, or
alternatively on the Bloch functions as follows:
\beq
\psid_\s(\br) = \sum_{\bR\nu} w^*_\nu(\br-\bR)\,\cd_{\bR\nu\s}
= \sum_{\bk\nu} \psiknu^*(\br)\,\cd_{\bk\nu\s}
\eeq
This leads to the following expression for the lattice part of the one-particle hamiltonian:
\beq
H_{0L} = \sum_{\bk\nu\s} \eknu\cdknus\cknus =
-\sum_{\bR\bR'}\sum_{\nu\s} \t^{(\nu)}_{\bR\bR'} c^\dagger_{\bR\nu\s}c_{\bR'\nu\s} +
\sum_{\bR\nu\s}\enu\,c^\dagger_{\bR\nu\s}c_{\bR\nu\s}
\eeq
with the hopping parameters and on-site energies given by:
\beq
\t^{(\nu)}_{\bR\bR'}=- \sum_\bk e^{i\bk\cdot(\bR-\bR')}\,\eknu =
-\int d\br\, w_\nu^*(\br-\bR)\,\left[-\frac{\hbar^2\nabla^2}{2m}+
V_L(\br)\right]\,w_\nu(\br-\bR')
\label{eq:hoppings}
\eeq
\beq
\enu=\sum_\bk\eknu
\eeq
Because the Bloch functions diagonalize the one-body hamiltonian, there are no inter-band
hopping terms in the Wannier representation considered here. Furthermore, for a separable
potential such as (\ref{eq:sep_potential}), close examination of (\ref{eq:hoppings}) show that
the oppings are only along the principal axis of the lattice: the hopping amplitudes
along diagonals vanish for a separable potential (see also Sec.~\ref{sec:frust}).

\begin{figure}
\begin{center}
\includegraphics[width=7 cm]{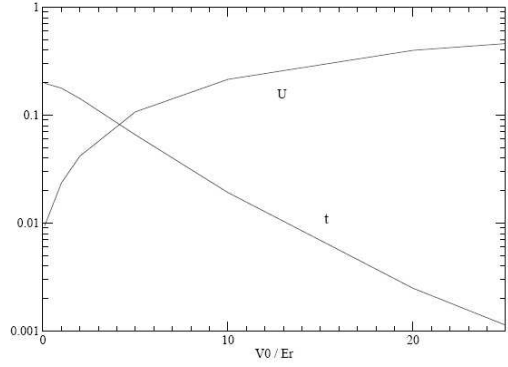}
\caption{Hopping amplitude $\t$ and on-site interaction energy $U$, as a function
of $V_0/E_R$, for the three-dimensional separable potential (\ref{eq:sep_potential})
corresponding to a cubic lattice. $\t$ is expressed in units of $E_R$ and
$U$ in units of $100\,E_R\,a_s/d$, with $a_s$ the scattering length and $d$ the lattice spacing.
From Ref~\cite{werner_dea}.}
\label{fig:U_t}
\end{center}
\end{figure}

Let us now turn to the interaction hamiltonian. The inter-particle distance
and lattice spacing are generally
much larger than the hard-core radius of the inter-atomic potential.
Hence, the details of the potential at short
distance do not matter. Long distance properties of the potential are characterized by the
{\it scattering length} $a_s$. As is well known, and described elsewhere in these lectures,
$a_s$ can be tuned over a wide range of positive or negative
values by varying the magnetic field close to a Feshbach resonance. Provided the extent of the
Wannier function is larger than the scattering length ($l_1\gg a_s$), the following
pseudopotential can be used:
\beq
V^{\s,-\s}_{\rm{int}}(\br-\br')=g\,\delta(\br-\br')\,\,\,,\,\,\,g\equiv\frac{4\pi\hbar^2a_s}{m}
\label{eq:pseudopot}
\eeq
The interaction hamiltonian then reads:
\beq
H_{\rm{int}} = g\,\int d\br\, \psid_\uparrow(\br)\psi_\uparrow(\br)
\psid_\downarrow(\br)\psi_\downarrow(\br)
\eeq
which can be written in the basis set of Wannier functions (assumed for simplicity to be
real) as follows:
\beq
H_{\rm{int}}=\sum_{\bR_1\bR_2\bR_3\bR_4}\sum_{\nu_1\nu_2\nu_3\nu_4}\,
U_{\bR_1\bR_2\bR_3\bR_4}^{\nu_1\nu_2\nu_3\nu_4}\,
\cd_{\bR_1\nu_1\uparrow}c_{\bR_2\nu_2\uparrow}
\cd_{\bR_3\nu_3\downarrow}c_{\bR_4\nu_4\downarrow}
\eeq
with:
\beq
U_{\bR_1\bR_2\bR_3\bR_4}^{\nu_1\nu_2\nu_3\nu_4}\,=g\,
\int d\br\, w_{\nu_1}(\br-\bR_1)w_{\nu_2}(\br-\bR_2)
w_{\nu_3}(\br-\bR_3)w_{\nu_4}(\br-\bR_4)
\eeq
The largest interaction term corresponds to two atoms on the same lattice site. Furthermore,
for a deep enough lattice, with less than two atoms per site on average, the second band is
well separated from the lowest one. Nelecting all other bands, and all interaction terms
except the largest on-site one, one obtains the single-band Hubbard model with a local
interaction term:
\beq
H_H\,=\,-\sum_{\bR\bR'\s} \t_{\bR\bR'} \cd_{\bR\s}c_{\bR'\s} +
U \sum_\bR \hat{n}_{\bR\uparrow}\hat{n}_{\bR\downarrow}
\eeq
with:
\beq
U\,=\,g\,\int d\br\, w_1(\br)^4
\label{eq:U_wannier}
\eeq
For a deep lattice, using the above estimate of the extension $l_1$ of the
Wannier function of the lowest band, this leads to~\cite{zwerger_mott_jopt_2003}
(compare to the hopping amplitude (\ref{eq:hop_deep}) which decays exponentially):
\beq
\frac{U}{E_R}\,\simeq\,\sqrt{\frac{8}{\pi}}\,a_sk_L\,\left(\frac{V_0}{E_R}\right)^{3/4}
\label{eq:U_deep}
\eeq
The hopping amplitude and on-site interaction strength $U$, calculated for the
lowest band of a three-dimensional separable potential, are plotted as a function of
$V_0/E_R$ in Fig.~\ref{fig:U_t}.

Let us finally discuss the conditions under which this derivation of a simple single-band
Hubbard model is indeed valid. We have made 3 assumptions: i) neglect the second band, ii) neglect other
interactions besides the Hubbard $U$ and iii) replace the actual interatomic potential
by the pseudopotential approximation. Assumption i) is justified provided the second band is
not populated (less than two fermions per site, and $V_0$ not too small so that the
two bands do not overlap, i.e $V_0 \gg 2.3 E_R$ cf. Fig.~\ref{fig:bandwidth}),
but {\it also} provided the energy cost for
adding a second atom on a given lattice site which already has one is indeed set by the interaction
energy. If $U$ as given by (\ref{eq:U_wannier}) becomes larger than
the separation $\Delta=\sum_\bk(\ve_{\bk 2}-\ve_{\bk 1})$
between the first two bands, then it is more favorable to add the second atom in the second
band (which then cannot be neglected, even if not populated). Hence one must have $U<\Delta$.
For the pseudopotential to be valid (assumption -iii), the typical distance between two atoms in a lattice well
(which is given by the extension of the Wannier function $l_1$) must be
larger than the scattering length: $l_1\gg a_s$.
Amusingly, for deep lattices, this actually coincides with the requirement $U\ll\Delta$
and boils down to (at large $V_0/E_R$):
\beq
\frac{a_s}{d}\,\lesssim\, \left(\frac{V_0}{E_R}\right)^{-1/4}
\label{eq:condition}
\eeq
In order to see this, one simply has to use the above estimates
of $l_1$ ($\sim d (E_R/V_0)^{1/4}$) and $U/E_R$ ($\sim a_s/d (V_0/E_R)^{3/4}$)
and that of the separation
$\Delta\simeq (E_R V_0)^{1/2}$ in this limit. Eq.~(\ref{eq:condition}) actually shows that
for a deep lattice, the scattering length should not be increased too much if one wants to
keep a Hubbard model with an interaction set by the scattering length itself and given by
(\ref{eq:U_deep}). For larger values of $a_s$, it may be that a one-band Hubbard description still
applies (see however below for the possible appearance of new interaction terms), but
with an effective $U$ given by the inter-band separation $\Delta$ rather than set by $a_s$. This
requires a more precise investigation of the specific case at hand~\footnote{This is reminiscent of the
so-called Mott insulator to charge-transfer insulator crossover in condensed matter physics}.

\begin{figure}
\begin{center}
\includegraphics[width=10 cm]{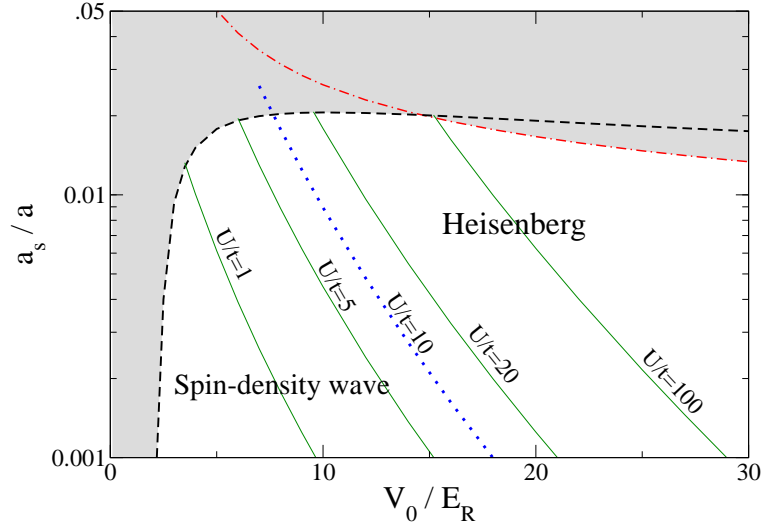}
\caption{Range of validity of the simple one-band Hubbard model description, for
a separable three-dimensional potential (\ref{eq:sep_potential}), as a function of lattice depth
(normalized to recoil energy) $V_0/E_R$, and scattering length (normalized to lattice spacing)
$a_s/d$.
In the shaded region, the one-band Hubbard description is questionable.
The dashed line corresponds to the condition $U/\Delta = 0.1$, with $\Delta$ the
gap to the second band: above this line,
other bands may have to be taken into account and the
pseudopotential approximation fails, so that $U$ is no longer given by
(\ref{eq:U_wannier}). The dashed-dotted line corresponds to $V_h/\t = 0.1$: above
this line, $V_h$ becomes sizeable. These conditions may be somewhat too restrictive,
but are meant to emphasize the points raised in the text.
Also indicated on the figure are: contour plots of the values of the Hubbard
coupling $U/\t$, and the regions corresponding
to the spin-density wave and Heisenberg regimes of the antiferromagnetic
ground-state at half-filling (Sec.\ref{sec:antiferro}).
The crossover between these regimes is indicated by the dotted line ($U/t=10$),
where $T_N/t$ is maximum. Figure from Ref.~\cite{werner_cooling_2005}.
}
\label{fig:hubbard}
\end{center}
\end{figure}

Finally, the possible existence of other interaction terms besides the on-site $U$ (-ii), and
when they can be neglected, requires a more careful examination.
These interactions must be smaller than $U$ but also than the hopping $\t$ which
we have kept in the hamiltonian.
In Ref.~\cite{werner_cooling_2005,werner_dea}, we considered this in more details
and concluded that the most `dangerous' coupling turns out to be a kind
of `density-assisted' hopping between
two nearest-neighbor sites, of the form:
\beq
V_h \sum_{\langle\bR\bR'\rangle}\sum_\s\, \hat{n}_{\bR,-\s}
\cd_{\bR\s}c_{\bR',\s} + \rm{h.c}
\eeq
with:
\beqno
V_h=g\,\int d\br\, w_1(\br)^3 w_1(\br+\mathbf{d}) =
g\,\left(\int dx w_x(x)^3w_x(x+d)\right)
\left(\int dy w_y(y)^4\right)
\left(\int dz w_z(z)^4\right)
\eeq
where $\mathbf{d}$ denotes a lattice translation between nearest-neighbor sites, and the
last formula holds for a separable potential.
The validity of the single-band Hubbard model also requires that $V_h \ll \t, U$.
All these requirements insuring that a simple Hubbard model description is valid
are summarized on Fig.~\ref{fig:hubbard}.

\section{The Mott phenomenon.}
\label{sec:mott}

Strong correlation effects appear when atoms ``hesitate'' between localized and
itinerant behaviour. In such a circumstance, one of the key difficulties is to describe
consistently an entity which is behaving simultaneously in a wave-like (delocalized) and
particle-like (localized) manner. Viewed from this perspective, strongly correlated quantum
systems raise questions which are at the heart of the quantum mechanical world.

The most dramatic example is the possibility of a phase transition
between two states: one in which atoms behave in an itinerant manner, and one
in which they are localized by the strong on-site repulsion in the potential wells
of a deep lattice. In the Mott insulating case, the energy gain which could be obtained by
tunneling between lattice sites ($\sim D\t \simeq W$) becomes unfavorable in comparison to the
cost of creating doubly occupied lattice sites ($\sim U$). This cost will have to be paid
for sure if there is, for example, one atom per lattice site on average.
This is the famous Mott transition. The proximity of a Mott insulating
phase is in fact responsible for many of the intriguing properties of strongly
correlated electron materials in condensed matter physics, as illustrated
below in more details. This is why the theoretical proposal~\cite{jaksch_lattice_prl_1998}
and experimental observation~\cite{greiner_mott_nature_2002}
of the Mott transition in a gas of ultra-cold bosonic atoms in an
optical lattice have truly been pioneering works establishing a bridge between
modern issues in condensed matter physics and ultra-cold atomic systems.

\subsection{Mean-field theory of the bosonic Hubbard model}
\label{sec:meanfield_bose}

Even though this school is devoted to fermions, I find it useful to briefly
describe the essentials of the mean-field theory of the Mott transition in
the bosonic Hubbard model. Indeed, this allows to focus on the key phenomenon (namely,
the blocking of tunneling by the on-site repulsive interaction) without having to
deal with the extra complexities of fermionic statistics and
spin degrees of freedom which complicate the issue in the case
of fermions (see below).

Consider the Hubbard model for single-component bosonic atoms:
\begin{equation}
H=-\sum_{ij} \t_{ij}\, b^\dagger_ib_j +
\frac{U}{2}\sum_i \hn_i(\hn_i-1)
-\mu \sum_i \hn_i
\end{equation}
As usually the case in statistical mechanics, a mean-field theory can be constructed
by replacing this hamiltonian on the lattice by an effective single-site
problem subject to a self-consistency condition. Here, this is naturally achieved by
factorizing the hopping term~\cite{fisher_bosehubbard_prb_1989,sheshadri_bosehubbard_epl_1993}:
$b^\dagger_ib_j \rightarrow \rm{const.} +
\langle b^\dagger_i\rangle b_j + b^\dagger_i \langle b_j\rangle + \rm{fluct.}$.
Another essentially equivalent formulation is based on the
Gutzwiller wavefunction~\cite{rokhsar_bosehubbard_prb_1991,krauth_bosehubbard_prb_1992}.
The effective 1-site hamiltonian for site $i$ reads::
\begin{equation}
h_{\rm{eff}}^{(i)}=
-\lambda_i b^\dagger -\lambda_i b +
\frac{U}{2} \hn(\hn-1)
-\mu \hn
\label{eq:singlesite_bose}
\end{equation}
In this expression, $\lambda_i$ is a ``Weiss field'' which is determined
self-consistently by the boson amplitude on the other sites of the lattice through the
condition:
\begin{equation}
\lambda_i = \sum_j \t_{ij}\, \langle b_j \rangle
\label{eq:scc_bose}
\end{equation}
For nearest-neighbour hopping on a uniform lattice of connectivity $z$, with all sites
being equivalent, this reads:
\begin{equation}
\lambda = z\,\t\,\langle b \rangle
\label{eq:scc_bose_uniform}
\end{equation}
These equations are easily solved numerically, by diagonalizing the effective
single-site hamiltonian (\ref{eq:singlesite_bose}), calculating $\bra b \ket$ and
iterating the procedure such that (\ref{eq:scc_bose_uniform}) is satisfied.
The boson amplitude $\bra b \ket$ is an order-parameter which is
non-zero in the superfluid phase. For densities corresponding to
an integer number $n$ of bosons per site on average, one finds that $\bra b \ket$ is
non-zero only when the coupling constant $U/\t$ is smaller than a critical
ratio $(U/\t)_c$ (which depends on the filling $n$). For $U/\t > (U/\t)_c$, $\bra b \ket$
(and $\lambda$) vanishes, signalling the onset of a non-superfluid phase in which
the bosons are localised on the lattice sites. For non-integer values of the density,
the system remains a superfluid for arbitrary couplings.

\begin{figure}
\begin{center}
\includegraphics[width=11cm]{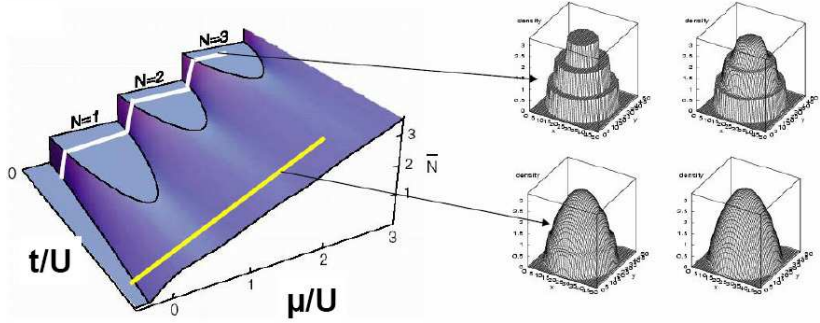}
\caption{Left: phase diagram of the Bose Hubbard model as a function of chemical potential
$\mu/U$ and coupling $\t/U$. An incompressible Mott insulator is found within each lobe of
integer density. Right: density profiles in a harmonic trap. The ``wedding cake'' structure
(see text) is
due to the incompressibility of the Mott insulator (numerical calculations courtesy of
H.Niemeyer and H.Monien, figure courtesy F.Gerbier).
}
\label{fig:phasediag_bose}
\end{center}
\end{figure}

It is instructive to analyze these mean-field equations close to the critical
value of the coupling: because $\lambda$ is then small, it can be treated in perturbation
theory in the effective hamiltonian (\ref{eq:singlesite_bose}). Let us start with
$\lambda=0$. We then have a collection of disconnected lattice sites (i.e no effective
hopping, often called the ``atomic limit'' in condensed matter physics).
The ground-state of an isolated site is the number state $|n\ket$ when the
chemical potential is in the range $\mu\in [(n-1)U,nU]$.
When $\lambda$ is small, the perturbed ground-state becomes:
\begin{equation}
|\psi_0\rangle = |n\rangle -
\lambda\,\left[\frac{\sqrt{n}}{U(n-1)-\mu}|n-1\rangle+
\frac{\sqrt{n+1}}{\mu-Un}|n+1\rangle\right]
\end{equation}
so that:
\begin{equation}
\langle\psi_0|b|\psi_0\rangle =
- \lambda\,\left[\frac{n}{U(n-1)-\mu}+
\frac{n+1}{\mu-Un}\right]
\end{equation}
Inserting this in the self-consistency condition yields:
\begin{equation}
\lambda = - z\,\t\,\lambda\,\left[\frac{n}{U(n-1)-\mu}+
\frac{n+1}{\mu-Un}\right]+\cdots
\end{equation}
where ``...'' denotes higher order terms in $\lambda$. As usual, the critical value of the
coupling corresponds to the vanishing of the coefficient of the term linear
in $\lambda$ (corresponding to the mass term of the expansion of the Landau free-energy).
Hence the critical boundary for a fixed average (integer) density $n$ is given by:
\begin{equation}
\frac{z\t}{U}\,=\,\frac{(n-\mu/U)(\mu/U-n+1)}{1+\mu/U}
\label{eq:critical_bose}
\end{equation}
This expression gives the location of the critical boundary as a function of the
chemical potential. In the ($\t/U,\mu/U$) plane, the phase diagram
(Fig.~\ref{fig:phasediag_bose})
consists of lobes inside which the density is integer and the system is a Mott insulator.
Outside these lobes, the system is a superfluid. The tip of a given lobe corresponds to the
the maximum value of the hopping at which an insulating state can be found. For
$n$ atoms per site, this is given by:
\begin{equation}
\frac{z\t}{U}|_{c,n}= \rm{Max}_{x\in[n-1,n]}
\frac{(n-x)[x-n+1]}{1+x} = \frac{1}{2n+1+2\sqrt{n(n+1)}}
\end{equation}
So that the critical interaction strength is
$(U/z\t)_c\simeq 5.8$ for $n=1$, and increases as $n$ increases ($(U/z\t)_c\sim 4n$ for large $n$).

\subsection{Incompressibility of the Mott phase and ``wedding-cake''
structure of the density profile in the trap}

The Mott insulator has a gap to density excitations and is therefore
an incompressible state: adding an extra particle costs a finite amount of energy.
This is clear from the mean-field calculation above: if we want to vary the average
density from infinitesimally below an integer value $n$ to infinitesimally above, we have
to change the chemical potential across the Mott gap:
\beq
\Delta_g(n) = \mu_{+}(n)-\mu_{-}(n)
\eeq
where $\mu_{\pm}$ are the solutions of the quadratic equation corresponding to
(\ref{eq:critical_bose}), i.e:
\beq
(\mu/U)^2-[2n-1-(z\t/U)] (\mu/U) + n(n-1)+(z\t/U) = 0
\eeq
yielding:
\beq
\Delta_g(n)=U\,\left[(\frac{z\t}{U})^2-2(2n+1)\frac{z\t}{U}+1\right]^{1/2}
\eeq
The Mott gap is $\sim U$ at large $U$ and vanishes at the critical coupling
($\propto\sqrt{U-U_c}$ within mean-field theory).

The existence of a gap means that the chemical potential can be changed within the
gap without changing the density. As a result, when the system is placed in a trap,
it displays density plateaus corresponding to the Mott state, leading to
a ``wedding cake'' structure of the density profile (Fig.~\ref{fig:phasediag_bose}).
This is easily
understood in the local density approximation, in which the local chemical potential
is given by:
$\mu(\br)=\bar{\mu}-m\omega_0^2 r^2/2$, yielding a maximum extension of the
plateau: $\sim (2\Delta_g/m\omega_0^2)^{1/2}$. Several authors have studied
these density plateaus beyond the LDA by numerical simulation (see e.g
\cite{batrouni_domains_prl_2002}), and they have been recently observed
experimentally~\cite{folling_shellstructure_prl_2006}.

\subsection{Fermionic Mott insulators and the Mott transition
in condensed matter physics}

The discussion of Mott physics in the fermionic case is somewhat complicated
by the presence of the spin degrees of freedom (corresponding e.g to 2 hyperfine
states in the context of cold atoms). Of course, we could consider single component
fermions, but two of those cannot be put on the same lattice site because of the
Pauli principle, hence spinless fermions with one atom per site on average simply
form a band insulator. Mott and charge density wave physics would show up in this context
when we have e.g one fermion out of two sites, but this requires inter-site
(e.g dipolar) interactions.

The basic physics underlying the Mott phenomenon in the case of two-component
fermions with one particle per site on average is the same as in the bosonic
case however: the strong on-site repulsion overcomes the kinetic energy
and makes it unfavorable for the particles to form an itinerant (metallic) state.
From the point of view of band theory, we would have a metal, with one atom per unit
cell and a half-filled band. Instead, at large enough values of $U/\t$, a Mott
insulating state with a charge gap develops.
This is purely charge physics, not spin physics.

One must however face the fact that
the naive Mott insulating state has a huge spin entropy: it is a paramagnet in
which the spin of the atom localized on a given site can point in either direction.
This huge degeneracy must be lifted as one cools down the system into its ground-state
(Nernst).
How this happens will depend on the details of the model and of the residual interactions
between the spin degrees of freedom.
In the simplest case of a two-component model on an unfrustrated (e.g. bipartite)
lattice, the spins order
into an {\it antiferromagnetic} ground-state. This is easily understood in strong coupling
$U\gg \t$ by Anderson's superexchange mechanism: in a single-band model, a nearest-neighbor
magnetic exchange is generated, which reads on each lattice bond:
\beq
J_{\rm{AF}}\,=\,\frac{4\t_{ij}^2}{U}
\eeq
This expression is easily understood from second-order degenerate perturbation theory
in the hopping, starting from the limit of decoupled sites ($\t=0$). Then, two given sites
have a 4-fold degenerate ground-state. For small $\t$, this degeneracy is lifted: the singlet
state is favoured because a high-energy virtual state is allowed in the perturbation expansion
(corresponding to a doubly occupied state), while no virtual excited state is connected to the
triplet state because of the Pauli principles (an atom with a given spin cannot hop to a site
on which another atom with the same spin already exists).
If we focus only on low-energies, much smaller than the gap to
density excitations ($\sim U$ at large $U$), we can consider the reduced Hilbert
space of states with exactly one particle per site. Within this low-energy Hilbert space,
the Hubbard model with one particle per site on average reduces to the quantum
Heisenberg model:
\beq
H_J\,=\,J_{\rm{AF}}\,\sum_{\langle ij\rangle}\,\mathbf{S}_i\cdot\mathbf{S}_j
\eeq
Hence, there is a clear {\it separation of scales} at strong coupling:
for temperatures/energies $T\lesssim U$, density fluctuations
are suppressed and the physics of a paramagnetic Mott insulator (with a large spin entropy)
sets in.
At a much lower scale $T\lesssim\Jaf$, the residual spin interactions set in and the true ground-state
of the system is eventually reached (corresponding, in the simplest case, to an ordered antiferromagnetic
state).

\begin{figure}
\begin{center}
\includegraphics[width=11cm]{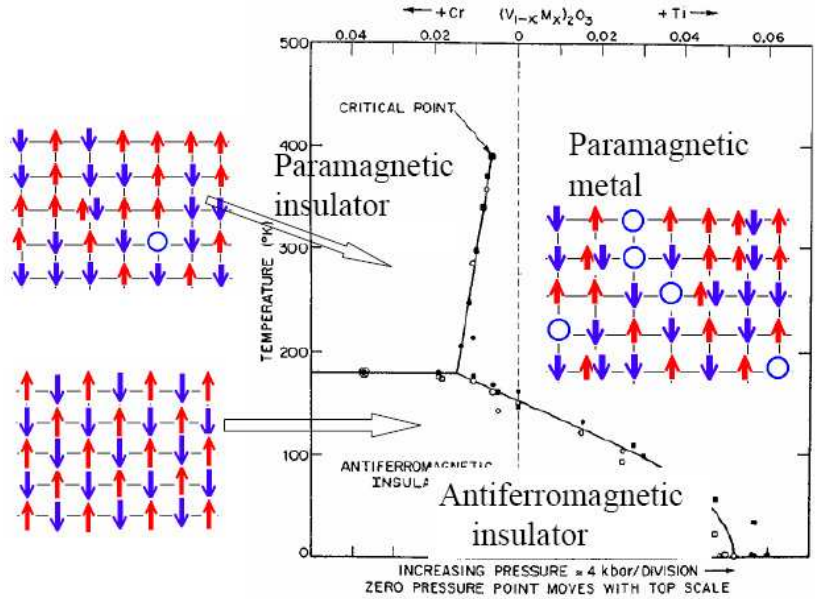}
\caption{Phase diagram of V$_2$O$_3$ as a function of pressure of Cr-substitution,
and temperature. The cartoons illustrate the nature of each phase (paramagnetic Mott
insulator, paramagnetic metal, antiferromagnetic Mott insulator).
}
\label{fig:v2o3}
\end{center}
\end{figure}

At this point, it is instructive to pause for a moment and ask what real materials do
in the condensed matter physics world. Materials with strong electronic correlations are
those in which the relevant electronic orbitals (those corresponding to energies close to the
Fermi energy) are quite strongly localized around the nuclei, so that a band theory description
in terms of Bloch waves is not fully adequate (and may even fail completely). This happens in
practice for materials containing partially filled $d$- and $f$-shells, such as transition
metals, transition-metal oxides, rare earths, actinides and their compounds, as well as
many organic conductors (which have small bandwidths).
In all these materials, Mott physics and the proximity to a Mott insulating
phase plays a key role.
In certain cases, these materials are poised rather close to the localisation/delocalisation transition
so that a small perturbation can tip off the balance. This is the case, for example, of a material
such as V$_2$O$_3$ (vanadium sesquioxide), whose phase diagram is displayed in
Fig.~\ref{fig:v2o3}.
The control parameter in this material is the applied pressure (or chemical substitution
by other atoms on vanadium sites), which change the unit-cell volume and hence the
bandwidth (as well, in fact, as other characteristics of the electronic structure, such as the
crystal-field splitting). It is seen from Fig.~\ref{fig:v2o3} that all three phases
discussed above are realized in this material.
At low pressure and high temperature, one has
a paramagnetic Mott insulator with fluctuating spins. As the pressure is increased, this
insulator evolves abruptly into a metallic state, through a first order transition line (which
ends at a critical endpoint at $T_c\simeq 450$K).
At low temperature $T<T_N\simeq 170\,$K, the paramagnetic Mott insulator orders
into an antiferromagnetic Mott insulator. Note that the characteristic temperatures
at which these transitions take place are considerably smaller than the bare
electronic energy scales ($\sim 1$eV$\simeq 12000\,$K).

On Fig.~\ref{fig:v2o3}, I have given for each phase a (much oversimplified) cartoon of what the
phase looks like in real space. The paramagnetic Mott insulator is a superposition of
essentially random spin configurations, with almost only
one electron per site and
very few holes and double occupancy. The antiferromagnetic insulator has N\'eel-like
long range order (but of course the wavefunction is a complicated object, not the
simple N\'eel classical wavefunction). The metal is the most complicated when looking at
it in real space: it is a superposition of configurations with singly occupied sites, holes,
and double occupancies.

Of course, such a material is far less controllable than ultra-cold atomic systems: as
we apply pressure many things change in the material, not only e.g the electronic
bandwidth. Also, not only the electrons are involved: increasing the lattice
spacing as pressure is reduced decreases the electronic cohesion of the crystal and the
ions of the lattice may want to take advantage of that to gain elastic energy: there is
indeed a discontinuous change of lattice spacing through the first-order Mott transition line.
Atomic
substitutions introduce furthermore some disorder into the material. Hence, ultra-cold atomic
systems offer an opportunity to disentangle the various phenomena and study these effects
in a much more controllable setting.

\subsection{(Dynamical) Mean-field theory for fermionic systems}

In section.~\ref{sec:meanfield_bose}, we saw how a very simple mean-field theory
of the Mott phenomenon can be constructed for bosons, by using $\langle b\rangle$ as
an order parameter of the superfluid phase and making an effective field (Weiss)
approximation for the inter-site hopping term. Unfortunately, this cannot be
immediately extended to fermions. Indeed, we cannot give an expectation value to
the single fermion operator, and $\langle c\rangle$ is not an order parameter of
the metallic phase anyhow.

A generalization of the mean-field concept to many-body fermion systems does
exist however, and is known as the ``dynamical mean-field theory'' (DMFT)
approach.
There are many review articles on the subject
(e.g \cite{georges_review_dmft,kotliar_dmft_physicstoday,georges_strong}),
so I will only describe
it very briefly here. The basic idea is still to replace the lattice system
by a single-site problem in a self-consistent effective bath. The exchange of
atoms between this single site and the effective bath is described by an amplitude,
or hybridization function~\footnote{Here, I use the Matsubara quantization formalism at finite
temperature, with $\omega_n=(2n+1)\pi/\beta$ and $\beta=1/kT$}, $\Delta(\iomn)$,
which is a function of energy (or time). It is a quantum-mechanical
generalization of the static Weiss field in classical statistical mechanics, and physically
describes the tendancy of an atom to leave the site and wander in the rest of the lattice.
In a metallic phase, we expect $\Delta(\omega)$ to be large at low-energy, while in the
Mott insulator, we expect it to vanish at low-energy so that motion to other sites is blocked.

The (site+effective bath) problem is described by an effective action, which for
the paramagnetic phase of the Hubbard model reads:
\beq
S_{\rm{eff}}=-\sum_n\sum_\s c^\dagger_{\s}(\iomn) [\iomn+\mu-\Delta(\iomn)] c_\s(\iomn)
+ U\,\int_0^\beta d\tau\,n_\uparrow\,n_\downarrow
\label{eq:seff_dmft}
\eeq
From this local effective action, a one-particle Green's function and self-energy can be
obtained as:
\beq
G(\tau-\tau') = - \langle T\,c_\s(\tau)c^\dagger_\s(\tau')\rangle_{\rm{eff}}
\eeq
\beq
\Sigma(\iomn)=\iomn+\mu-\Delta(\iomn)-G(\iomn)^{-1}
\eeq
The self-consistency condition, which closes the set of dynamical mean-field theory
equations, states that the Green's function and self-energy of the (single-site+bath)
problem coincides with the corresponding local (on-site) quantities in the original
lattice model. This yields:
\beq
G(\iomn)=\sum_\bk\,\frac{1}{\iomn+\mu-\Sigma(\iomn)-\ek}=
\sum_\bk\,\frac{1}{\Delta(\iomn)+G(\iomn)^{-1}-\ek}
\label{eq:scc_dmft}
\eeq
Equations (\ref{eq:seff_dmft},\ref{eq:scc_dmft}) form a set of two equations which determine self-consistently
both the local Green's function $G$ and the dynamical Weiss field $\Delta$. Numerical methods are
necessary to solve these equations, since one has to calculate the Green's function of a
many-body (albeit local) problem. Fortunately, there are several computational algorithms
which can be used for this purpose.

\begin{figure}
\begin{center}
\includegraphics[width=8cm]{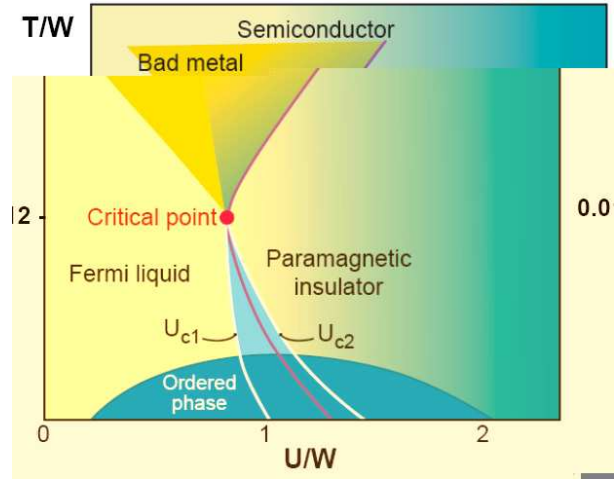}
\caption{Schematic phase diagram of the half-filled fermionic Hubbard model, as
obtained from DMFT. It is depicted here for the case of a frustrated lattice (e.g with
next-nearest neighbour hopping), which reduces the transition temperature into phases with
long-range spin ordering. Then, a first-order transition from a metal to a paramagnetic
Mott insulator becomes apparent. For the unfrustrated case, see next section. Adapted from
\cite{kotliar_mott_science_2003}.
}
\label{fig:dmft_phasediag}
\end{center}
\end{figure}

On Fig.~\ref{fig:dmft_phasediag}, I display the schematic shape of the generic
phase diagram obtained with dynamical mean-field theory, for the one band Hubbard model with
one particle per site. At high temperature, there is a crossover from a Fermi liquid (metallic)
state at weak coupling to a paramagnetic Mott insulator at strong coupling. Below some critical
temperature $T_c$, this crossover turns into a first-order transition line. Note that $T_c$ is a
very low energy scale: $T_c\simeq W/80$, almost two orders of magnitude smaller than the bandwidth.
Whether this critical temperature associated with the Mott transition
can be actually reached depends on the concrete model under consideration. In the simplest
case, i.e for a single band with nearest-neighbor hopping on an unfrustrated lattice,
long range antiferromagnetic spin ordering takes place already at a temperature
far above $T_c$, as studied in more details in the next section. Hence, only a finite-temperature
crossover, not a true phase transition,
into a paramagnetic Mott insulator will be seen in this case. However, if antiferromagnetism
becomes frustrated, the N\'eel temperature can be strongly suppressed, revealing
genuine Mott physics, as shown in the schematic phase diagram of Fig.~\ref{fig:dmft_phasediag}.

\section{Ground-state of the 2-component Mott insulator: antiferromagnetism.}
\label{sec:antiferro}

\begin{figure}
\begin{center}
\includegraphics[width=9cm]{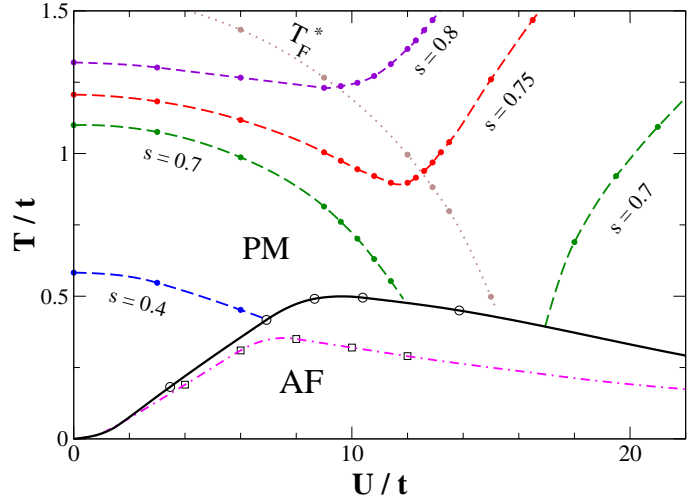}
\caption{Phase diagram of the half-filled Hubbard model on the cubic lattice:
antiferromagnetic (AF) and paramagnetic (PM) phases. Transition temperature
within the DMFT approximation (plain curve, open circles) and from the QMC calculation of
Ref.~\cite{staudt_af_hubbard_epjb_2000} (dot-dashed curve, squares).
Dashed lines: isentropic curves (s=0.4,0.7,0.75,0.8), computed within DMFT.
Dotted line: quasiparticle coherence scale $T_F^*(U)$. See
Ref.~\cite{werner_cooling_2005} for more details.
}
\label{fig:T_N and isentropics}
\end{center}
\end{figure}

Here, I consider in more details the simplest possible case of a one-band
Hubbard model, with nearest-neighbor hopping on a bipartite (e.g cubic) lattice and
one atom per site on average. The phase diagram, as determined by various methods
(Quantum Monte Carlo, as well as the DMFT approximation) is displayed on
Fig.~\ref{fig:T_N and isentropics}. There are only two phases: a high-temperature
paramagnetic phase, and a low-temperature antiferromagnetic phase which is an insulator
with a charge gap. Naturally, within the high-temperature phase, a gradual crossover from itinerant
to Mott localized is observed as the coupling $U/\t$ is increased, or as the temperature is
decreased below the Mott gap ($\sim U$ at large $U/\t$). Note that the mean-field estimate
of the Mott critical temperature $T_c\simeq W/80$ is roughly a factor of two lower than
that of the maximum value of the N\'eel temperature for this model ($\sim W/40$), so we do not
expect the first-order Mott transition line and critical endpoint to be apparent
in this unfrustrated situation.

Both the weak coupling and strong coupling sides of the phase diagram are
rather easy to understand. At weak coupling, we can treat $U/\t$ by a Hartree-Fock
decoupling, and construct a static mean-field theory of the antiferromagnetic
transition. The broken symmetry into $(A,B)$ sublattices reduces the Brillouin zone
to half of its original value, and two bands are formed which read:
\beq
E_\bk^\pm = \pm\, \sqrt{\ek^2+\Delta_g^2/4}
\eeq
In this expression, $\Delta$ is the Mott gap, which within this Hartree approximation is
directly related to the staggered magnetization of the ground-state
$m_s=\langle n_{A\uparrow}-n_{A\downarrow}\rangle=
\langle n_{B\downarrow}-n_{B\uparrow}\rangle$ by:
\beq
\Delta_g = U\, m_s
\eeq
This leads to a self-consistent equation for the gap (or staggered magnetization):
\beq
\frac{U}{2}\sum_{\bk\in\rm{RBZ}}\frac{1}{\sqrt{\ek^2+\Delta_g^2/4}}=1
\eeq
At weak-coupling, where this Hartree approximation is a reasonable starting point,
the antiferromagnetic instability occurs for arbitrary small $U/\t$ and the gap, staggered
magnetization and N\'eel temperature are all exponentially small. In this regime, the antiferromagnetism is
a ``spin density-wave'' with wavevector $\mathbf{Q}=(\pi,\cdots,\pi)$ and a very weak
modulation of the order parameter.

It should be noted that this spin-density wave mean-field theory provides a band theory
(Slater) description of the insulating ground-state: because translational
invariance is broken in the antiferromagnetic ground-state, the Brillouin zone is halved, and the
ground-state amounts to fully occupy the lowest Hartree-Fock band. This is because there is
no separation of energy scales at weak coupling: the spin and charge degrees of freedom get frozen
at the same energy scale. The existence of a band-like description in the weak coupling
limit is often a source of confusion, leading some people to overlook that Mott physics is
primarily a charge phenomenon, as it becomes clear at intermediate and strong coupling.

In the opposite regime of strong coupling $U\gg\t$, we have already seen that the Hubbard model
reduces to the Heisenberg model at low energy. In this regime, the N\'eel temperature is
proportional to $\Jaf$, with quantum fluctuations significantly reducing $T_N/\Jaf$ from its
mean-field value: numerical simulations~\cite{staudt_af_hubbard_epjb_2000}
yield $T_N/\Jaf \simeq 0.957$ on the cubic lattice. Hence, $T_N/\t$ becomes small
(as $\sim \t/U$) in strong coupling. In between these two regimes, $T_N$ reaches a
maximum value (Fig.~\ref{fig:T_N and isentropics}).

On Fig.~\ref{fig:hubbard}, we have indicated
the two regimes corresponding to spin-density wave and Heisenberg antiferromagnetism,
in the $(V_0/E_R,a_s/d)$ plane. In fact, the crossover between these two regimes is
directly equivalent to the BCS-BEC crossover for an attractive interaction. For one particle per
site, and a bipartite lattice, the Hubbard model with $U>0$ maps onto the same model
with $U<0$ under the particle-hole transformation (on only one spin species):
\beq
c_{i\uparrow}\,\rightarrow\,\,\widetilde{c}_{i\uparrow}\,\,\,,\,\,\,
c_{i\downarrow}\,\rightarrow\,(-1)^i\,\widetilde{c}^\dagger_{i\downarrow}
\eeq
with $(-1)^i=+1$ on the A-sublattice and $=-1$ on the B-sublattice.
The spin density wave (weak coupling) regime corresponds to the BCS one and
the Heisenberg (strong-coupling) regime to the BEC one.

\section{Adiabatic cooling: entropy as a thermometer.}
\label{sec:cool}

As discussed above, the N\'eel ordering temperature is a rather low
scale as compared to the bandwidth. Considering the value of
$T_N$ at maximum and taking into account the appropriate range of $V_0/E_R$ and
the constraints on the Hubbard model description, one would estimate that
temperatures on the scale of $\sim 10^{-2} E_R$ must be reached.
This is at first sight a bit deceptive, and
one might conclude that the prospects for cooling down to low enough temperatures
to reach the antiferromagnetic Mott insulator are not so promising.

In Ref.~\cite{werner_cooling_2005} however, we have argued that one should
in fact think in terms of {\it entropy rather than temperature}, and that
interaction effects in the optical lattice lead to adiabatic cooling mechanisms
which should help.

\begin{figure}
\begin{center}
\includegraphics[width=7 cm]{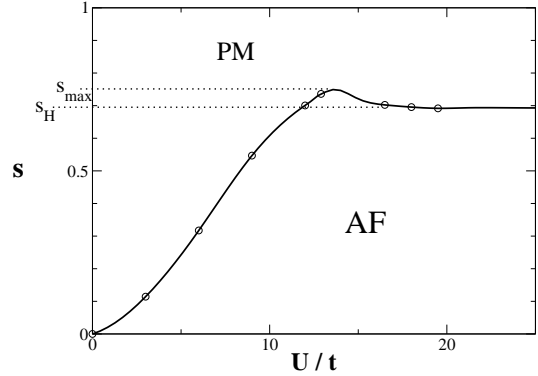}
\caption{Schematic phase diagram of the one-band Hubbard model at half filling,
as a function of entropy and coupling constant. The marked dots are from
a DMFT calculation (in which case
$s_H=\ln 2$), but the shape of the critical boundary is expected to be general
(with $s_H<\ln 2$ reduced by quantum fluctuations).}
\label{fig:s_N}
\end{center}
\end{figure}

Consider the entropy per particle of the homogeneous half-filled Hubbard
model: this is a function $s(T,U)$ of the temperature and coupling~\footnote{The entropy
depends only on the ratios $T/\t$ and $U/\t$: here we express for simplicity the temperature
and coupling strength in units of the hopping amplitude $\t$.}. The entropy itself
is a good thermometer since it is an increasing function of temperature
($\partial s/\partial T > 0$).
Assuming that an adiabatic process is possible, the key point to reach the AF
phase is to be able to prepare the system in a state which has a smaller entropy than
the entropy at the N\'eel transition, i.e along the critical boundary:
\beq
s_N(U)\,\equiv\,s\left(T_N(U),U\right)
\eeq
It is instructive to think of the behaviour of this quantity as a function of
$U$.
At weak-coupling (spin-density wave regime),
$s_N(U)$ is expected to be exponentially small. In contrast, in the opposite Heisenberg regime at
large $U/\t$, $s_N$ will reach a finite value $s_H$, which is the entropy of the quantum Heisenberg
model at its critical point. $s_H$ is a pure number which depends only on the specific lattice
of interest. Mean-field theory of the Heisenberg model yields $s_H=\ln 2$, but quantum
fluctuations will reduce this number. In \cite{werner_cooling_2005}, this reduction was
estimated to be of order $50\%$ on the cubic lattice, i.e $s_H\simeq \ln 2/2$, but a precise
numerical calculation would certainly be welcome.
How does $s_N$ evolve from weak to strong coupling~? A rather general argument suggests that
it should go through a maximum $\smax >s_H$. In order to see this, we take a derivative
of $s_N(U)$ with respect to coupling, observing that:
\begin{equation}
\frac{\partial s}{\partial U} = - \frac{\partial \pd}{\partial T}
\label{eq:partial s}
\end{equation}
In this expression, $\pd$ is the probability that a given site is doubly occupied:
$\pd\equiv\langle n_{i\uparrow} n_{i\downarrow} \rangle$. This relation stems from the
relation between entropy and free-energy: $s= -\partial f/\partial T$ and $\partial f/\partial U=\pd$
Hence, one obtains:
\begin{equation}
\frac{ds_N}{dU} =
\frac{c(T_N)}{T_N}\,\frac{dT_N}{dU} - \frac{\partial\pd}{\partial T}|_{T=T_N}
\label{eq:delsU}
\end{equation}
in which $c(T,U)$ is the specific heat per particle: $c=T\partial s/\partial T$.
If only the first term was present in the r.h.s of this equation, it would imply
that $s_N$ is maximum exactly at the value of the coupling where $T_N$ is
maximum (note that
$c(T_N)$ is finite ($\alpha<0$)
for the 3D-Heisenberg model).
Properties of the double occupancy discussed below show that
the second term in the r.h.s has a similar variation than the first one.
These considerations
suggest that $s_N(U)$ does reach a maximum
value $s_{\rm{max}}$ at intermediate coupling, in the same range of $U$ where $T_N$ reaches a maximum.
Hence, $s_N(U)$ has the general form sketched on Fig.~\ref{fig:s_N}.
This figure can be viewed as a phase diagram of the half-filled Hubbard model, in
which entropy itself is used as a thermometer, a very natural
representation when addressing adiabatic cooling.
Experimentally, one may first cool down the gas (in the absence of the optical
lattice) down to a temperature where the entropy per particle is lower than
$s_{\rm{max}}$ (this corresponds to $T/T_F<s_{\rm{max}}/\pi^2$ for a trapped ideal gas).
Then, by branching on the optical lattice adiabatically, one could increase
$U/t$ until one particle per site is reached over most of the trap: this should allow
to reach the antiferromagnetic phase. Assuming that the timescale for
adiabaticity is simply set by the hopping, we observe that typically
$\hbar/\t\sim 1 \textrm{ms}$.

The shape of the isentropic curves in the plane $(U/\t,T/\t)$, represented
on Fig.~\ref{fig:T_N and isentropics}, can also be discussed on
the basis of Eq.~(\ref{eq:delsU}). Taking a derivative of the equation defining
the isentropic curves:
$s(T_i(U),U)=\rm{const.}$, one obtains:
\begin{equation}
c(T_i)\,\frac{\partial T_i}{\partial U} =
T_i\,\frac{\partial\pd}{\partial T}|_{T=T_i}
\label{eq:isentropics}
\end{equation}
The temperature-dependence of the probability of
double occupancy $\pd(T)$ has been studied in details
using DMFT (i.e in the mean-field limit of large dimensions).
When $U/\t$ is not too large, the double
occupancy first {\it decreases} as temperature is increased from $T=0$
(indicating a higher degree of localisation), and
then turns around and grows again.
This apparently counter-intuitive behavior is a direct
analogue of the Pomeranchuk effect in liquid Helium 3: since the
(spin-) entropy is larger in a localised state than when the fermions
form a Fermi-liquid (in which $s\propto T$), it is favorable to
increase the degree of localisation upon heating. The minimum of
$\pd(T)$ essentially coincides with the {\it quasiparticle coherence scale}
$\tstar(U)$: the scale below which coherent (i.e long-lived) quasiparticles
exist and Fermi liquid theory applies (see section.~\ref{sec:quasi}).
Mott localisation implies that $\tstar(U)$ is a rapidly decreasing function of $U/\t$
(see Fig.~\ref{fig:T_N and isentropics}).  The ``Pomeranchuk cooling'' phenomenon therefore
applies only as long as $\tstar>T_N$, and hence when $U/\t$ is not too
large. For large $U/\t$, Mott localisation dominates for all
temperatures $T<U$ and suppresses this effect.
Since $\partial\pd/\partial T<0$ for $T<\tstar(U)$ while
$\partial\pd/\partial T>0$ for $T>\tstar(U)$, Eq.(\ref{eq:isentropics})
implies that the isentropic curves of the half-filled Hubbard
model (for not too high values of the entropy) must have a
negative slope at weak to
intermediate coupling, before turning around at stronger coupling, as
shown on Fig.~\ref{fig:T_N and isentropics}.

It is clear from the results of Fig.~\ref{fig:T_N and isentropics}
that, starting from a low enough initial value of the entropy per site,
adiabatic cooling can be achieved by
either increasing $U/\t$ starting from a small value, or decreasing $U/\t$ starting from
a large value (the latter requires however to cool down the gas
while the lattice is already present).
We emphasize that this cooling mechanism is an interaction-driven
phenomenon: indeed, as $U/\t$ is increased, it allows to lower the
{\it reduced temperature} $T/t$, normalized to the natural scale
for the Fermi energy in the presence of the lattice. Hence, cooling is not
simply due here to the tunneling amplitude $\t$ becoming
smaller as the lattice is turned on, which is the effect for non-interacting fermions
discussed in Ref.~\cite{blakie_cooling_fermions} and Sec.~\ref{sec:scales} above.
At weak coupling and low temperature,
the cooling mechanism can be related to the effective mass of
quasiparticles ($\propto 1/\tstar$) becoming heavier as $U/\t$ is increased,
due to Mott localisation. Indeed, in this regime,
the entropy is proportional to $T/\tstar(U)$. Hence,
conserving the entropy while increasing $U/\t$ adiabatically from
$(U/\t)_i$ to $(U/\t)_f$ will reduce the final temperature in comparison to
the initial one $T_i$ according to: $T_f/T_i=\tstar(U_f)/\tstar(U_i)$.

This discussion is based on the mean-field behaviour of the probability of double
occupancy $\pd(T,U)$. Recently~\cite{dare_adiabatic_2006}, a direct study in three dimensions confirmed the
possibility of ``Pomeranchuk cooling'', albeit with a somewhat reduced efficiency as compared to
mean-field estimates. In two dimensions however, this effect is not expected to apply,
due to the rapid growth of antiferromagnetic correlations which quench the spin entropy.
A final note is that the effect of the trapping potential has not been taken into account in this
discussion, and further investigation of this effect in a trap would certainly be worthwile.

\section{The key role of frustration.}
\label{sec:frust}

In the previous section, we have seen that, for an optical lattice without geometrical
frustration (e.g a bipartite lattice with nearest-neighbour hopping amplitudes), the
ground-state of the half-filled Hubbard model is a Mott insulator with long-range
antiferromagnetic spin ordering. Mott physics has to do with the blocking of density
(charge) fluctuations however, and spin ordering is just a consequence. It would be nice to be able
to emphasize Mott physics by getting rid of the spin ordering, or at least reduce the temperature
scale for spin ordering. One way to achieve this is by {\it geometrical frustration} of the
lattice, i.e having next-nearest neighbor hoppings ($\t^\prime$) as well. Indeed, such a hopping will
induce a next-nearest neighbor antiferromagnetic superexchange, which obviously leads to
a frustrating effect for the antiferromagnetic arrangement of spins on each triangular plaquette
of the lattice.

\begin{figure}
\begin{center}
\includegraphics[width=7 cm]{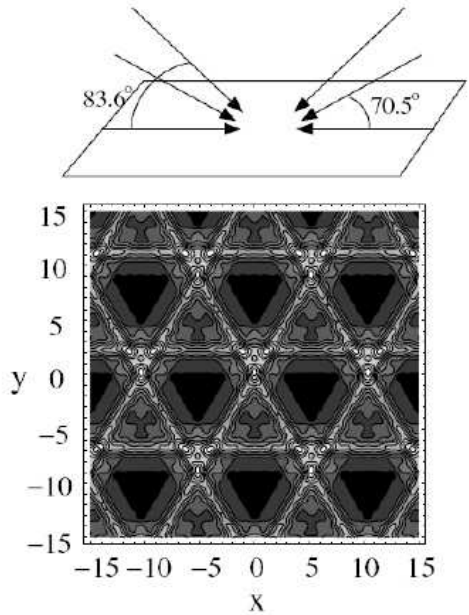}
\caption{Laser setup (top) proposed in Ref.~\cite{santos_kagome_prl_2004} to realize a
trimerized kagome optical lattice (bottom). Figure adapted from \cite{santos_kagome_prl_2004}.}
\label{fig:kagome}
\end{center}
\end{figure}

It is immediately seen that inducing next nearest-neighbour hopping along a diagonal
link of the lattice requires
a non-separable optical potential however. Indeed, in a separable potential, the Wannier
functions are products over each coordinate axis: $W(\br-\bR)=\prod_{i=1}^D w_i(r_i-R_i)$.
The matrix elements of the kinetic energy $\sum_i \hbar^2\nabla_i^2/2m$ between two Wannier
functions centered at next-nearest neighbor sites along a diagonal link thus vanish
because of the orthogonality of the $w_i$'s between nearest neighbors.
Engineering the optical potential such as to obtain a desired set of tight-binding parameters
is an interesting issue which I shall not discuss in details in these notes however.
A classic reference on this subject is the detailed paper by Petsas
et al.~\cite{petsas_optical_lattices_pra_1994}. Recently, Santos et
al.~\cite{santos_kagome_prl_2004} demonstrated the possibility of generating a
``trimerized'' Kagome lattice, a highly frustrated two-dimensional lattice, with a
tunable ratio of the intra-triangle to inter-triangle exchange (Fig.~\ref{fig:kagome}).

\subsection{Frustration can reveal ``genuine'' Mott physics}
\label{sec:genuine_mott}

As mentioned above, frustration can help revealing Mott physics by pushing
spin ordering to lower temperatures. One of the possible consequences is the
appearance of a genuine (first-order) phase transition at finite temperature
between a metallic (itinerant)
phase at smaller $U/\t$ and a paramagnetic Mott insulating phase at large $U/\t$, as
depicted in Fig.~\ref{fig:dmft_phasediag}. Such a transition is indeed found within
dynamical mean-field theory (DMFT), i.e in the limit of large lattice connectivity, for
frustrated lattices.
A first-order transition is observed in real materials as well (e.g
in V$_2$O$_3$, cf. Fig.~\ref{fig:v2o3}) but in this case the lattice degrees of freedom
also participate (although the transition is indeed electronically driven).
There are theoretical indications that, in the presence of frustration, a first
order Mott transition at finite temperature exists for a rigid lattice beyond mean-field
theory (see e.g~\cite{parcollet_cdmft_finiteTMott_prl_2004}), but no solid proof either.
In solid-state physics, it is not possible to
suppress the coupling of electronic instabilities
to lattice degrees of freedom, hence the experimental demonstration of
this is hardly possible.
This is a question that ultra-cold atomic systems might help answering.

The first-order transition line
ends at a second-order critical endpoint: there is indeed no symmetry distinction
between a metal and an insulator at finite temperature and it is logical that one
can then find a continuous path from one to the other around the critical point.
The situation is similar to the liquid-gas transition, and in fact it is expected
that this phase transition is in the same universality class: that of the Ising model
(this has been experimentally demonstrated for V$_2$O$_3$~\cite{limelette_v2o3_science}).
A qualitative analogy with the liquid-gas transition can actually been
drawn here: the Mott insulating phase has very few doubly occupied, or empty, sites
(cf. the cartoons in Fig.~\ref{fig:v2o3}) and hence corresponds to a low-density or
gas phase (for double occupancies), while the metallic phase has many of them and
corresponds to the higher-density liquid phase.

\begin{figure}
\begin{center}
\includegraphics[width=6 cm]{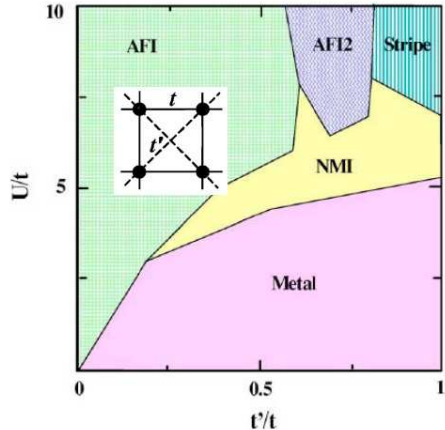}
\caption{Ground-state phase diagram of the two-dimensional Hubbard model with
nearest-neighbor and next nearest-neighbor hopping, as obtained in
Ref.~\cite{mizusaki_spinliquid_prb_2006} from the ``path-integral renormalization group method''.
A non-magnetic Mott insulator (NMI) is stabilized for large enough frustration $\t^\prime/\t$ and
intermediate coupling $U/\t$.
A similar model with n.n.n hopping along only one of the diagonals (anisotropic triangular lattice)
was studied in Ref.~\cite{kyung_organics_prl_2006} using a cluster extension of DMFT, and an additional
d-wave superconducting phase was found in this study.
}
\label{fig:phasediag_ttprime}
\end{center}
\end{figure}

One can also ask whether it is possible to stabilize a paramagnetic Mott phase
as the {\it ground-state}, i.e suppress spin ordering down to $T=0$. Several recent studies of
frustrated two-dimensional models found this to happen at intermediate coupling $U/\t$ and for large enough
frustration $\t^\prime/\t$, with non-magnetic insulating and possibly d-wave superconducting ground states
arising (Fig.~\ref{fig:phasediag_ttprime}).

\subsection{Frustration can lead to exotic quantum magnetism}
\label{sec:exotic_mag}

The above question of suppressing magnetic ordering down to $T=0$ due to frustration can also
be asked in a more radical manner by considering the strong-coupling limit
$U/\t\rightarrow\infty$. There, charge (density) fluctuations are entirely suppressed and
the Hubbard model reduces to a quantum Heisenberg model. The question is then whether
quantum fluctuations of the spin degrees of freedom only, can lead to a ground-state without
long-range order. Studying this issue for frustrated Heisenberg models
or related models has been a very active field of theoretical condensed matter physics for the past
20 years or so, and I simply direct
the reader to existing reviews on the subject, e.g Ref.~\cite{misguich_review,lhuillier_review_2005}.
Possible disordered phases are {\it valence bond crystals}, in which translational
symmetry is broken and the ground-state can be qualitatively thought of as a specific
paving of the lattice by singlets living on bonds. Another, more exotic, possibility
is that the ground-state can be thought of as a resonant superposition of singlets (a
sort of giant benzene molecule): this is the {\it ``resonating valence bond''} idea proposed in
the pioneering work of Anderson and Fazekas. There are a few examples of this, one
candidate being the Heisenberg model on the kagome lattice (Fig.~\ref{fig:kagome}).
Naturally, obtaining such unconventional states in ultra-cold atomic systems, and more
importantly being able to measure the spin-spin correlations and excitation spectrum
experimentally would be fascinating.

One last remark in this respect, which establishes an interesting connection between
exotic quantum magnetism and Bose condensation. A spin-1/2 quantum
Heisenberg model with a ground-state which is not ordered and does not break
translational symmetry (e.g a resonating valence bond ground-state) is
analogous, in a precise formal sense, to a specific interacting model of
hard-core bosons which would {\it remain a normal liquid} (not a crystal,
not a superfluid) down to $T=0$. Hence, somewhat ironically, an unconventional ground-state means,
in the context of quantum magnetism, {\it preventing Bose condensation}.
To see this, we observe that a quantum spin-1/2 can be represented with a
hard-core boson operator $b_i$ as:
\beqno
S^+_i=b^\dagger_i\,\,\,,\,\,\,
S^-_i=b_i\,\,\,,\,\,\,
S^z_i=b^\dagger_ib_i-\frac{1}{2}
\eeq
with the constraint that at most one boson can live on a given site
$b^\dagger_ib_i = 0,1$
(infinite hard-core repulsion).
The anisotropic Heisenberg (XXZ) model then reads:
\beqno
H = J_\perp\sum_{\bra ij\ket} [b^\dagger_i b_j +b^\dagger_j b_i] +
J_z \sum_{\bra ij\ket} (b^\dagger_i b_i-1/2)(b^\dagger_j b_j-1/2)
\eeq
Hence, it is an infinite-U bosonic Hubbard model with an additional interaction
between nearest-neighbor sites (note that dipolar interactions can
generate
those for real bosons). The superfluid phase for the bosons
correspond to a phase with XY-order in the spin language,
a crystalline (density-wave) phase with broken translational symmetry
to a phase with antiferromagnetic ordering of the $S^z$ components, and
a normal Bose fluid to a phase without any of these kinds of orders.

\section{Quasiparticle excitations in strongly correlated fermion systems,
and how to measure them.}
\label{sec:quasi}

\subsection{Response functions and their relation to the
spectrum of excitations}

Perhaps even more important than the nature of the ground-state of
a many-body system is to understand the nature of the excited states,
and particularly of the {\it low-energy} excited states (i.e close to the
ground-state). Those are
the states which control the response of the system to a weak perturbation,
which is what we do when we perform a measurement without disturbing the
system too far out of equilibrium~\footnote{Ultra cold atomic systems, as already
stated in the introduction, also offer the possibility of performing measurements far
from equilibrium quite easily, which is another -fascinating- story.}.
When the perturbation is weak, linear response theory can be used, and
in the end what is measured is the correlation function of some observable
(i.e of some operator $\hO$):
\beq
C_O(\br,\br';t,t')\,=\,\bra \hO(\br,t)\,\hO^\dagger(\br^\prime,t^\prime)\ket
\eeq
In this expression, the operators evolve in the Heisenberg representation, and the
brackets denote either an average in the ground-state (many-body) wave function (for
a measurement at $T=0$) or, at finite temperature, a thermally weighted average with
the equilibrium Boltzmann weight. How the behaviour of this correlation function is
controlled by the spectrum of excited states is easily understood by inserting a complete
set of states in the above expression (in order to make the time evolution explicit)
and obtaining the following spectral representation (given here at $T=0$ for simplicity):
\beq
C_O(\br,\br';t,t')\,=\,\sum_n e^{-\frac{i}{\hbar}(E_n-E_0)(t-t^\prime)}\,
\bra\Phi_0|\hO(\br)|\Phi_n\ket\bra\Phi_n|\hO(\br^\prime)^\dagger|\Phi_0\ket
\eeq
In this expression, $\Phi_0$ is the ground-state (many-body) wave function,
and the summation
is over all admissible many-body excited states (i.e having non-zero matrix elements).

A key issue in the study of ultra-cold atomic systems is to devise measurement techniques
in order to probe the nature of these many-body states. In many cases, one can resort to
spectroscopic techniques, quite similar in spirit to what is done in condensed matter
physics. This is the case, for example, when the observable $\hO$ we want to access
is a local observable such as the local density or the local spin-density. Light (possibly
polarized) directly couples to those, and light scattering is obviously the tool of choice
in the context of cold atomic systems. Bragg scattering~\cite{stamperkurn_bragg_prl_1999}
can be indeed used to measure the
density-density dynamical correlation function $\bra\rho(\br,t)\rho(\br',t')\ket$ and
polarized light also allows one to probe~\cite{carusotto_bragg_jpb_2006} the spin-spin response
$\bra\mathbf{S}(\br,t)\mathbf{S}'(\br',t')\ket$. In condensed matter physics, analogous
measurements can be done by light or neutron scattering.

One point is worth emphasizing here, for condensed matter physicists. In
condensed matter physics, we are used to thinking of visible or infra-red light (not X-ray !)
as a {\it zero-momentum} probe, because the wavelength is much bigger than inter-atomic distances.
This is not the case for atoms in optical lattices ! For those, the lattice spacing is set by
the wavelength of the laser, hence lasers in the same range of wavelength can be used to
sample the momentum-dependence of various observables, with momentum transfers
possibly spanning the full extent of the
Brillouin zone.

Other innovative measurement techniques of various two-particle
correlation functions have recently been proposed and
experimentally demonstrated in the
context of ultra-cold atomic systems, some of which are reviewed
elsewhere in this set of lectures, e.g noise correlation
measurements~\cite{altman_noise_pra_2004,folling_noise_nature_2005,greiner_noise_prl_2005},
or periodic modulations of the
lattice~\cite{stoferle_modulation_prl_2004,kollath_modulation_prl_2006}.

The simplest examples we have just discussed involve {\it two-particle} correlation functions
(density-density, spin-spin), and hence
probe at low energy the spectrum of particle-hole excitations, i.e excited states
$\Phi_n$ which are coupled to the ground-state via an operator conserving particle number.
In contrast, one may want to probe {\it one-particle correlation functions}, which
probe excited states of
the many-body system with one atom added to it, or one atom removed, i.e coupled
to the ground-state via a single particle process. Such a correlation
function (also called the single-particle Green's function $G_1$) reads:
\beq
\bra T_t\psi(\br,t)\psi^\dagger\,(\br^\prime,t^\prime)\ket \equiv i\,G_1(\br,\br';t,t')
\eeq
in which $T_t$ denotes time ordering. The corresponding spectral decomposition involves the
{\it one-particle spectral function} (written here, for simplicity, for a homogeneous
system -so that crystal momentum is a good quantum number- and at $T=0$):
\begin{eqnarray}\nonumber
A(\bk,\omega)=&\sum_n |\langle
\Phi_n^{N-1}|c_{\,\bk}|\Phi_0^N\rangle|^2\,
\delta(\omega+\mu+E_n-E_0)\,\,\,(\omega<0)\\
=&\sum_n |\langle
\Phi_n^{N+1}|c_{\,\bk}^\dagger|\Phi_0^N\rangle|^2\,
\delta(\omega+\mu+E_0-E_n)\,\,\,(\omega>0)
\end{eqnarray}
The spectral function is normalized to unity for each momentum, due to the
anticommutation of fermionic operators:
\beq
\int_{-\infty}^{+\infty} A(\bk,\omega)\, d\omega = 1
\eeq

As explained in the next section, the momentum and frequency dependence
of this quantity contains key information about the important low-energy
excitations of fermionic systems (hole-like, i,e corresponding to
the removal of
one atom, for $\omega<0$, and particle-like
for $\omega>0$). Let us note that,
for Bose systems with a finite condensate density $n_0$,
the two-particle correlators are closely related to the one-particle
correlators via terms such as $n_0\,\langle\psi^\dagger(\br,
t)\,\psi(\br',t')\rangle$. By contrast, in Fermi systems the
distinction between one- and two-particle correlators is
essential.

A particular case is the equal-time correlator,
$\bra\psi^\dagger(\br,t)\psi(\br',t)\ket$, i.e the one-body density
matrix, whose Fourier transform is the momentum distribution in the
ground-state:
\beq
N(\bk) = \bra\Phi_0|\cd_\bk\,c_\bk|\Phi_0\ket = \int_{-\infty}^0 A(\bk,\omega) d\omega
\eeq
For ultra-cold atoms, this can be measured in {\it time of flight} experiments.
Conversely, rf-spectroscopy experiments~\cite{chin_rf_science_2004} give some access
to the frequency dependence of the one-particle spectral function, but not to its
momentum dependence.

\subsection{Measuring one-particle excitations by stimulated Raman scattering}

In condensed matter physics,
angle-resolved photoemission spectroscopy (ARPES)
provides a direct probe of the one-particle
spectral function (for a pedagogical introduction,
see~\cite{damascelli_ARPESintro_physscripta_2004}). This technique
has played a key role in revaling the highly unconventional
nature of single-particle excitations in cuprate
superconductors~\cite{damascelli_rmp_2003}. It consists in measuring the
energy and momentum of electrons emitted out of the solid exposed
to an incident photon beam. In the simplest approximation, the
emitted intensity is directly proportional to the single-electron spectral
function (multiplied by the Fermi function and by some matrix elements).

\begin{figure}
\begin{center}
\includegraphics[width=7 cm]{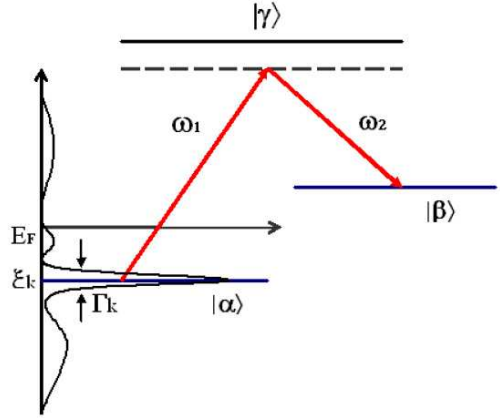}
\caption{\label{fig:raman}
Raman process: transfer from an internal state $\alpha$ to another internal state
$\beta$ through an excited state $\gamma$. The
momentum-resolved spectral function is schematized, consisting of a quasiparticle peak
and an incoherent background. From Ref.~\cite{dao_raman}.
}
\end{center}
\end{figure}

In Ref.~\cite{dao_raman}, it was recently proposed to use
stimulated Raman spectroscopy as a
probe of one-particle excitations, and of the frequency and
momentum dependence of the spectral function, in a two-component mixture of
ultracold fermionic atoms in two internal states $\a$ and $\a'$.
Stimulated Raman spectroscopy has been considered previously in
the context of cold atomic gases, both as an outcoupling technique
to produce an atom laser~\cite{hagley_atomlaser_science_1999} and
as a measurement technique for
bosons~\cite{japha_stimulated_prl_1999,luxat_tunneling_pra_2002,
blakie_raman_2005,mazets_raman_prl_2005} and
fermions~\cite{torma_raman_prl_2000,yi_raman_2006}. In the Raman process of
Fig.~\ref{fig:raman}, atoms are transferred from $\a$ into
another internal state $\b\neq\a,\a'$, through an intermediate
excited state $\g$, using two laser beams of wavectors
$\bk_{1,2}$ and frequencies $\omega_{1,2}$. If $\omega_1$ is
sufficiently far from single photon resonance to the excited $\g$
state, we can neglect spontaneous emission.
The total transfer rate to state $\b$ can be
calculated~\cite{japha_stimulated_prl_1999,luxat_tunneling_pra_2002,blakie_raman_2005}
using the Fermi golden rule and eliminating the excited state:
\beqno
R(\Dk,\Dom)=
|C|^{2}n_{1}(n_{2}+1)\int_{-\infty}^{\infty}\,dt \int
\!d\br\,d\br'\,e^{i[\,\Dom\,t
-\Dk\cdot(\br-\br')]}\, g_\b(\br,\br';t)
\langle \psi^{+}_{\alpha}(\br,t)\psi_{\alpha}(\br',0) \rangle
\label{eq:rate1}
\eeq
Here $\Dk=\bk_1-\bk_2$ and $\Dom=\omega_1-\omega_2+\mu$ with $\mu$
the chemical potential of the interacting gas, and $n_{1,2}$
the photon numbers present in the laser beams. Assuming that no
atoms are initially present in $\b$ and that the scattered atoms
in $\b$ do not interact with the atoms in the initial $\a,\a'$
states, the free propagator for $\b$-state atoms in vacuum is to
be taken: $g_\b(\br,\br';t)\equiv \langle
0_{\b}|\,\psib(\br,t)\psidb(\br',0)|0_{\b}\rangle$.
For a uniform system,
the transfer rate can be
related to the spectral function $A(\bk,\omega)$ of atoms in the
internal state $\a$ by~\cite{luxat_tunneling_pra_2002}:
\begin{equation}
R(\Dk,\Dom)\propto\,
\int \! d\bk\, n_{F}\,(\varepsilon_{\bk\beta}-\Dom)\,
A(\bk-\Dk,\varepsilon_{\bk\beta}-\Dom)
\label{eq:rate2}
\end{equation}
in which the Green's function has been expressed in terms of
the spectral function and the Fermi factor $n_F$.
In the presence of a trap, the confining potential can be treated in the
local density approximation by integrating the above expression
over the radial coordinate, with a position-dependent chemical potential.
From (\ref{eq:rate2}), the similarities and differences with ARPES are clear:
in both cases, an atom is effectively removed from the interacting gas, and
the signal probes the spectral function. In the case of ARPES, it is directly
proportional to it, while here an additional momentum integration is involved
if the atoms in state $\beta$ remain in the trap. One the other hand, in the
present context, one can in principle vary the momentum transfer $\Dk$ and regain
momentum resolution in this manner. Alternatively,
one can cut off the trap and perform a time of flight experiment~\cite{dao_raman},
in which case the measured rate is directly proportional to
$n_{F}\,(\varepsilon_{\bk\beta}-\Dom)\,
A(\bk-\Dk,\varepsilon_{\bk\beta}-\Dom)$, in closer analogy to ARPES. Varying the
frequency shift $\Dom$ then allows to sample different regions of the Brillouin
zone~\cite{dao_raman}.

\subsection{Excitations in interacting Fermi systems: a crash course}

Most interacting fermion systems have low-energy excitations which are well-described
by ``Fermi liquid theory'' which is a low-energy effective theory of these excitations.
In this description, the low-energy excitations are built out of
{\it quasiparticles}, long-lived
(coherent) entities carrying the same quantum numbers than the original particles.
There are three key quantities characterizing the quasiparticle excitations:
\begin{itemize}
\item Their dispersion relation, i.e the energy $\xik$
(measured from the ground-state energy) necessary to create such
an excitation with (quasi-) momentum $\bk$. The interacting system
possesses a Fermi surface (FS) defined by the location in
momentum space on which the excitation energy vanishes:
$\xi_{\kF}=0$. Close to a given point on the FS, the
quasiparticle energy vanishes as:
$\xik\sim\vF(\kF)\cdot(\bk-\kF)+\cdots$, with $\vF$ the local
Fermi velocity at that given point of the Fermi surface.

\item The spectral weight $Z_\bk\leq 1$ carried by these quasiparticle
excitations, in comparison to the total spectral weight ($=1$, see above) of
all one-particle excited states of arbitrary energy and fixed momentum.

\item Their lifetime $\Gk^{-1}$. It is finite away from the Fermi surface,
as well as at finite temperature. The quasiparticle lifetime diverges however at $T=0$ as
$\bk$ gets close to the Fermi surface. Within Fermi liquid theory, this happens in a specific manner
(for phase-space reasons), as $\Gk\sim\xik^2$. This insures the overall coherence of the
description in terms of quasiparticles,
since their inverse lifetime vanishes faster than their energy.

\end{itemize}

Typical signatures of strong correlations are the following effects (not necessarily
occurring simultaneously in a given system): i) strongly renormalized Fermi velocities, as
compared to the non-interacting (band) value, corresponding e.g to a large interaction-induced
enhancement of the effective mass of the quasiparticles, ii) a strongly suppressed quasiparticle
spectral weight $Z_\bk\ll 1$, possibly non-uniform along the Fermi surface, iii) short quasiparticle
lifetimes. These strong deviations from the non-interacting system can sometimes be
considerable: the ``heavy fermion'' materials for example (rare-earth compounds) have quasiparticle
effective masses which are several hundred times bigger than the mass from band theory, and in spite of
this are mostly well described by Fermi liquid theory.

The quasiparticle description applies only at low energy, below some characteristic energy
(and temperature) scale $\tstar$, the quasiparticle coherence scale.
Close to the Fermi surface, the one-particle spectral function displays a clear
separation of energy scales, with a sharp coherent peak carrying spectral weight
$\zk$ corresponding to
quasiparticles (a peak well-resolved in energy means long-lived excitations), and
an ``incoherent'' background carrying spectral weight $1-\zk$. A convenient form to
have in mind (Fig.~\ref{fig:raman}) is:
\beq
A(\bk,\omega)\,\simeq\, \zk\,\frac{\Gk}{\pi[(\omega-\xik)^2+\Gk^2]}\,+\,A_{\rm{inc}}(\bk,\omega)
\eeq
Hence, measuring the spectral function, and most notably the evolution of the quasiparticle
peak as the momentum is swept through the Fermi surface, allows one to probe the key properties
of the quasiparticle excitations: their dispersion (position of the peak),
lifetime (width of the peak) and spectral weight (normalized to the incoherent background,
when possible), as well of course as the location of the Fermi surface of the
interacting system in the Brillouin zone. In \cite{dao_raman}, it was shown that the shape
of the Fermi surface, as well as some of the quasiparticle properties can be
determined, in the cold atoms context, from the Raman spectroscopy described above.
For the pioneering experimental determination of Fermi surfaces in weakly or non-interacting fermionic
gases in optical lattices, see~\cite{kohl_fermisurface_prl_2005}.

What about the ``incoherent'' part of the spectrum (which in a strongly correlated
system may carry most of the spectral weight...) ? Close to the Mott transition, we expect
at least one kind of well-defined high energy excitations to show up in this incoherent spectrum. These are
the excitations which consist in removing a particle from a site which
is already occupied, or adding a particle on such a site. The energy difference separating these
two excitations is precisely the Hubbard interaction $U$. These excitations, which are
easier to think about in a local picture in real-space (in contrast to the wave-like, quasiparticle
excitations), form two broad dispersing peaks in the spectral functions: the so-called
Hubbard ``bands''.

In the mean-field (DMFT) description of interacting fermions and of the Mott transition,
the quasiparticle weight $Z$ is uniform along the Fermi surface. Close to the Mott transition,
$Z$ vanishes and the effective mass ($m^\star/m=1/Z$ in this theory)
of quasiparticles diverges. The quasiparticle coherence scale is $\tstar\simeq Z\,T_F$, with
$T_F$ the Fermi energy ($\sim$ bandwidth) of the non-interacting system: this coherence scale
also becomes very small close to the transition, and Hubbard bands carry most of the
spectral weight in this regime.

\subsection{Elusive quasiparticles and nodal-antinodal dichotomy:
the puzzles of cuprate superconductors}

The cuprate superconductors, which are quasi two-dimensional doped
Mott insulators, raise some fundamental questions about the description of
excitations in strongly interacting fermion systems. In the ``normal'' (i.e
non-superconducting) state of these materials, strong departure
from Fermi liquid theory is observed. Most notably, at doping levels smaller
than the optimal doping (where the superconducting $T_c$ is maximum), i.e
in the so-called ``underdoped'' regime:
\begin{itemize}
\item Reasonably well-defined quasiparticles are only observed close to the
diagonals of the Brillouin zone, i.e close to the ``nodal points'' of the Fermi
surface where the superconducting gap vanishes. Even there, the lifetimes are shorter and
appear to have a different energy and temperature dependence than that of Fermi liquid
theory.
\item In the opposite regions of the Fermi surface (``antinodal'' regions),
the spectral function shows no sign of a quasiparticle pleak in the normal state. Instead,
a very broad lineshape is found in ARPES, whose leading edge is {\it not centered at $\omega=0$},
but rather at a finite energy scale. The spectral function appears to have its maximum away
from the Fermi surface, i.e the density of low-energy excitations is strongly depleted
at low-energy: this is the ``pseudo-gap'' phenomenon. The pseudo-gap shows up in many other
kinds of measurements in the under-doped regime.
\end{itemize}
\begin{figure}
\begin{center}
\includegraphics[width=6 cm]{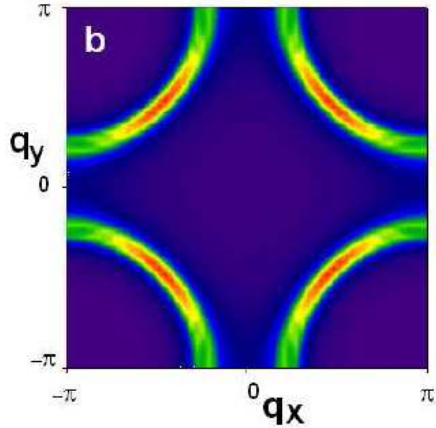}
\caption{Illustration of the dichotomy between ``nodal'' and ``antinodal''
regions of the Fermi surface, as observed in cuprate superconductors.
Colour coding corresponds to increasing intensity of the quasiparticle peak. Such effects
could be revealed in cold atomic systems by stimulated Raman spectroscopy measurements,
as proposed in Ref.~\cite{dao_raman}.}
\label{fig:nodal}
\end{center}
\end{figure}
Hence, there is a strong dichotomy between the nodal and antinodal regions in the normal state.
The origin of this dichotomy is one of the key issues in the field. One possibility is
that the pseudo-gap is due to a hidden form of long-range order which competes with
superconductivity and is responsible for suppressing excitations except in nodal
regions. Another possibility is that, because of the proximity to the Mott transition in
such low-dimensional systems, the quasiparticle coherence scale (and most likely
also the quasiparticle weight) varies strongly along the Fermi surface, hence suppressing
quasiparticles in regions where the coherence scale is smaller than temperature.

This nodal-antinodal dichotomy is illustrated in Fig.~\ref{fig:nodal}.
This figure has actually been obtained from a
simulated intensity plot of the Raman rate (\ref{eq:rate2}),
using a phenomenological form of the spectral function appropriate for cuprates.
It is meant to illustrate how future experiments on ultra-cold fermionic atoms in
two-dimensional optical lattices might be able to address some of the fundamental issues in the
physics of strongly correlated quantum systems.

\acknowledgments
I am grateful to Christophe Salomon, Massimo Inguscio and Wolfgang Ketterle
for the opportunity to lecture at the wonderful Varenna school on
``Ultracold Fermi Gases'',
to Jean Dalibard and Christophe Salomon
at the Laboratoire Kastler-Brossel
of Ecole Normale Sup\'erieure for stimulating my interest in this field and for collaborations,
and to Massimo Capone, Iacopo Carusotto, Tung-Lam Dao, Syed Hassan, Olivier Parcollet and Felix Werner
for collaborations related to the topics of these lectures. I also acknowledge useful discussions with
Immanuel Bloch, Frederic Chevy, Eugene Demler, Tilman Esslinger and Thierry Giamarchi.
My work is supported by the Centre National de la Recherche Scientifique, by Ecole
Polytechnique and by the Agence Nationale de la Recherche under contract
``GASCOR''.

%

\providecommand{\bysame}{\leavevmode\hbox to3em{\hrulefill}\thinspace}
\providecommand{\MR}{\relax\ifhmode\unskip\space\fi MR }
\providecommand{\MRhref}[2]{%
  \href{http://www.ams.org/mathscinet-getitem?mr=#1}{#2}
}
\providecommand{\href}[2]{#2}

\end{document}